\documentclass[aps,prb,twocolumn,flotfix,amsmath,amssymb,superscriptaddress]{revtex4-2}
\usepackage[english]{babel}
\usepackage{amsmath}
\usepackage{bm}
\usepackage{graphicx,bbm} 
\usepackage{times}
\usepackage{epsfig} 
\usepackage{soul}
\usepackage[utf8]{inputenc}
\usepackage[colorlinks,linkcolor=blue,citecolor=blue,urlcolor=blue]{hyperref}
\usepackage{color}
\usepackage{latexsym}
\usepackage{ulem}
\definecolor{nred} {RGB}{224,0,0}
\definecolor{nblue} {RGB}{28,130,185}
\definecolor{dgreen} {RGB}{78,138,21}

\begin{document} 

\title{Coexistence of diffusive and ballistic transport in integrable quantum lattice models}
\author{P. Prelov\v{s}ek}
\affiliation{Jo\v{z}ef Stefan Institute, SI-1000 Ljubljana, Slovenia}
\affiliation{Faculty of Mathematics and Physics, University of Ljubljana, SI-1000 
Ljubljana, Slovenia}
\author{M. Mierzejewski}
\affiliation{Department of Theoretical Physics, Faculty of Fundamental Problems of 
Technology, Wroc\l aw University of Science and Technology, 50-370 Wroc\l{a}w, Poland}
\author{J. Herbrych}
\affiliation{Department of Theoretical Physics, Faculty of Fundamental Problems of 
Technology, Wroc\l aw University of Science and Technology, 50-370 Wroc\l{a}w, Poland}

\date{\today}
\begin{abstract}
We investigate the high-temperature dynamical conductivity $\sigma(\omega)$ in two 
one-dimensional integrable quantum lattice models: the anisotropic XXZ spin chain and 
the Hubbard chain. The emphasis is on the metallic regime of both models, where 
besides the ballistic component, the regular part of conductivity might reveal a 
diffusive-like transport. To resolve the low-frequency dynamics, we upgrade the 
microcanonical Lanczos method enabling studies of finite-size systems with up to 
$L\leq 32$ sites for XXZ spin model with the frequency resolution 
$\delta \omega \sim 10^{-3} J$. Results for the XXZ chain reveal a fine structure of 
$\sigma(\omega)$ spectra, which originates from the discontinuous variation of the 
stiffness, previously found at commensurate values of the anisotropy parameter 
$\Delta$. Still, we do not find a clear evidence for a diffusive component, at least 
not for commensurate values of $\Delta$, particularly for $\Delta =0.5$, as well as for 
$\Delta \to 0$. Similar is the conclusion for the 
Hubbard model away from half-filling, where the spectra reveal more universal 
behavior.
\end{abstract}

\maketitle

\section {Introduction} 
One of the basic features of the integrable quantum many-body lattice is the 
possibility of the ballistic/dissipationless transport at finite temperatures (for a 
recent review see \cite{bertini21}). This property, which is manifested in a finite 
value of corresponding transport stiffnesses $D>0$, has been well established in the 
most investigated one-dimensional (1D) integrable model, the anisotropic XXZ spin 
chain within the easy-plane regime with the anisotropy $\Delta <1$, but also in the 
1D Hubbard model away from half-filling. Finite stiffness $D>0$ has been resolved via 
the sensitivity of levels to the imposed magnetic flux \cite{castella95}, using the 
thermodynamic Bethe Ansatz (TBA) \cite{zotos99,benz05,pavlis20,urichuk21}, and with more 
rigorous bounds via the Mazur inequality \cite{zotos97} which relate $D$ to the 
overlap with local and quasilocal conserved quantities 
\cite{prosen11,prosen13,prosen14,pereira14}. The latter result agree also with a more general 
approach via the generalized Hydrodynamics (GHD) 
\cite{ilievski17,ilievski171,bulchandani18}. Furthermore, results for $D>0$ and 
ballistic transport have been confirmed in numerous numerical studies of finite XXZ 
chains \cite{zotos96,castella96,naef98,hmeisner03,hmeisner07,rigol08,znidaric11,herbrych11,steinigeweg14,steinigeweg15,karrasch15,karrasch17}. 
In spite of these advances there remains an open question whether analytical theories 
also quantitatively fix values of $D$, in particular its dependence on the anisotropy 
$\Delta$ within the XXZ chain in the high-$T$ regime 
\cite{prosen13,steinigeweg14,karrasch15,ljubotina17,sanchez17,bulchandani18,mierzejewski21} 
(see the discussion in \cite{bertini21}).

Much less attention has been devoted to the whole dynamical response, as represented 
by the real part of the frequency-dependent conductivity at $T>0$, 
\begin{equation}
\sigma(\omega) = 2 \pi D\delta(\omega)+ \sigma_{\mathrm{\mathrm{reg}}} (\omega)\,,
\label{sigw}
\end{equation}
which can be (in a metallic regime of considered models) decomposed into the 
ballistic $D>0$ component and the regular (incoherent) part 
$\sigma_{\mathrm{reg}}(\omega)$. Exact-diagonalization (ED) results on finite-size 
XXZ chains \cite{zotos96,herbrych11}, as well as on the particular case of the 
Hubbard chain \cite{castella96}, indicate on vanishing (dc) limit, 
$\sigma_{\mathrm{reg}}^0 = \sigma_{\mathrm{reg}}(\omega \to 0) \to 0$, consistent with the 
argument based on the level crossing in integrable lattice models \cite{herbrych12} 
implying $\sigma_{\mathrm{reg}}(\omega \to 0) \propto\omega^2$, at least for finite-size 
systems. Less conclusive are 
results obtained via time-dependent density-matrix renormalization group (tDMRG) 
method on larger system but with restricted time evolution (or equivalently with 
limited frequency resolution), allowing for $\sigma^0_{\mathrm{reg}} >0$ 
\cite{karrasch15,karrasch17}. The latter can be interpreted as coexistence of 
ballistic transport and (subleading) diffusion response. This question is challenging
since some analytical approaches \cite{sirker09,sirker11}, and in particular more recently 
the GHD approach 
\cite{denardis18,ilievski18,gopalakrishnan18,agrawal20,bulchandani21}, generally 
predict besides $D >0$ also $\sigma_{\mathrm{reg}}^0 >0$. I.e., within the 
XXZ chain the GHD yields finite values of $\sigma_{\mathrm{reg}}^0 >0$ at the 
commensurate values $\Delta_m =\cos(\pi/m)$ and, moreover, singular 
$\sigma_{\mathrm{reg}}(\omega \to 0) \propto \omega^{-\alpha}, \alpha>0$ 
\cite{agrawal20,bulchandani21} behavior.
 
To comment on the dynamical transport $\sigma(\omega)$, at least from the perspective 
of the linear response in finite-size systems with periodic boundary conditions (PBC), 
we perform the numerical calculation 
of high-$T$ limit of $\tilde \sigma(\omega)=T \sigma(\omega)$ in XXZ model, scanning 
the whole range of anisotropies $\Delta<1$, but also in the Hubbard chain away from 
half-filling. Note that both models exhibit the finite stiffness ${\cal D}=T D >0$ 
for the considered model parameters. To resolve the low-$\omega$ regime, we employ, 
besides ED for smaller systems, the upgraded microcanonical Lanczos method (MCLM) 
\cite{long03} with the high-$\omega$ resolution, i.e., for the XXZ chain of length 
$L\leq 32$ we reach $\delta \omega \sim 10^{-3} J$ (equivalent to time evolution up 
to $\tau \sim 5.10^3/J$). Such a resolution allows to disentangle, according to 
Eq.~\eqref{sigw}, well enough the dissipationless part from the low-$\omega$ 
$\tilde \sigma_{\mathrm{reg}}(\omega)$. 

\begin{figure*}[tb]
\includegraphics[width=1.0\textwidth]{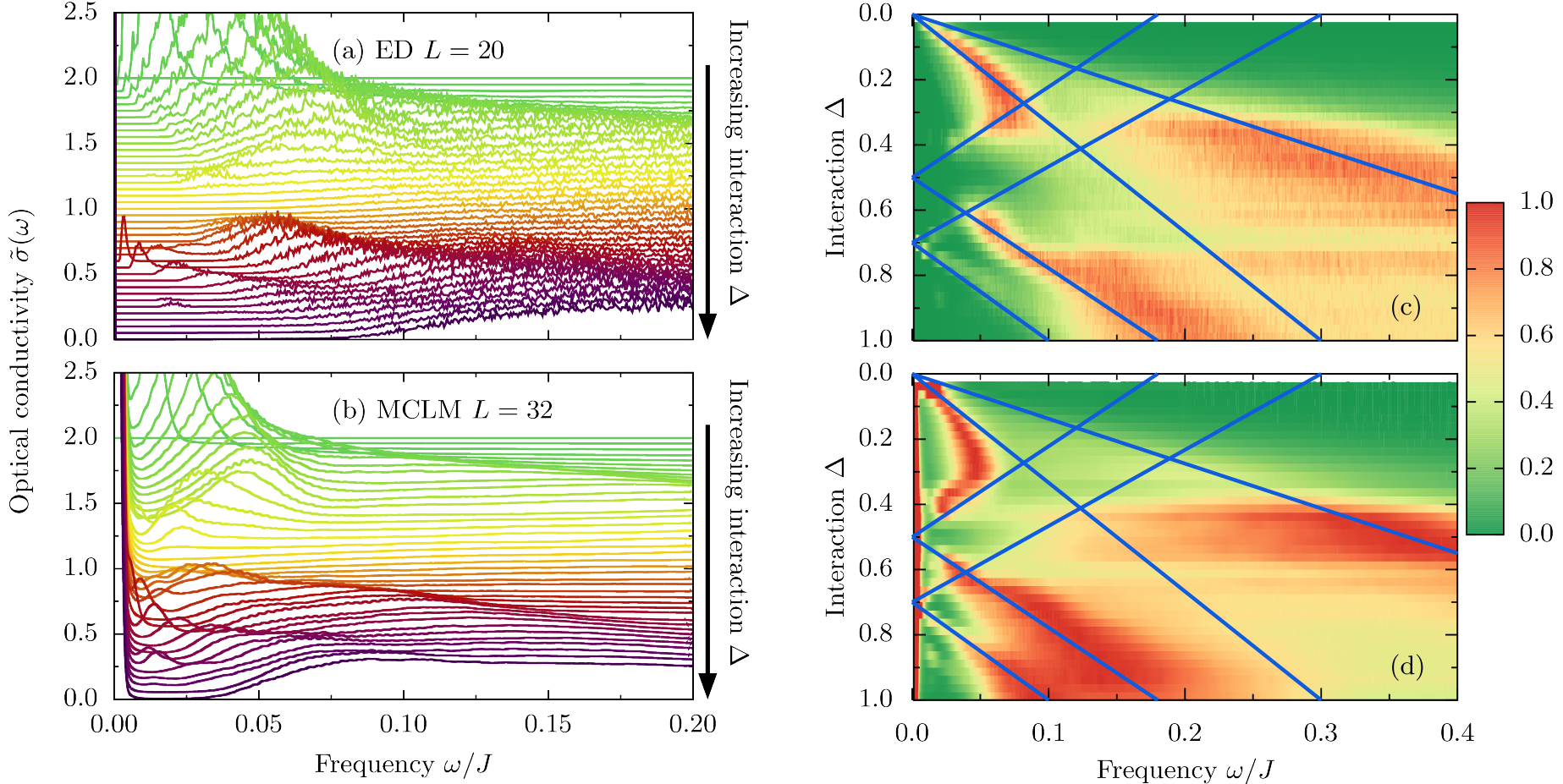}
\caption{High-$T$ spin conductivity 
$\tilde{\sigma}(\omega)=T\sigma_{\mathrm{reg}}(\omega)$ in the XXZ spin chain for the 
anisotropy range $\Delta \leq 1$ (\mbox{$\Delta=0,0.025,\dots,1.0$}), as calculated by (a) ED 
for $L=20$ sites and (b) MCLM for $L=32$ sites. Each consecutive curve is given an offset for 
better visibility. (c) and (d) show heatmaps of results from (a) and (b), 
respectively. For clarity, the latter results are normalized by maximum of 
$\tilde \sigma_{\mathrm{reg}}(\omega)$. Guidelines on both panels mark the positions 
of the low-frequency peaks obtained for $L=20$. }
\label{fig1}
\end{figure*}

In Fig.~\ref{fig1} we present one of the main results of our findings (in more detail 
discussed in Sec.~\ref{sec:stru}), i.e., changing the value of anisotropy $\Delta$ 
reveal a quite complicated fine-structure of $\tilde \sigma_{\mathrm{reg}}(\omega)$ 
spectra, here presented as the result of full ED for $L=20$ system, as well as for 
$L=32$ chain evaluated with MCLM. The structure can be traced back to discontinuities 
(or at least anomalies) of ${\cal D}$ found at commensurate $\Delta=\Delta_m$ 
\cite{bertini21} (even in finite systems). The spectral weight related with the 
discontinuities, 
\mbox{${\cal D}(\Delta_m)-\lim_{\Delta \to \Delta_m} {\cal D}(\Delta)$}, is 
transferred to low-frequency peaks of $\tilde \sigma_{\mathrm{reg}} (\omega)$ 
centered at $\omega_p \propto |\Delta-\Delta_m |$. Still, we do not find a clear 
evidence (or at least very small upper bound) for $\tilde \sigma_{\mathrm{reg}}^0$ at 
commensurate $\Delta_m = \cos(\pi/m)$, in particular for $\Delta=\Delta_3=0.5$. In 
order to identify the positions of peaks at $\omega_p$, in Fig.~\ref{fig1}(c,d) we 
plot $\tilde \sigma_{\mathrm{reg}} (\omega)/\tilde \sigma_{\mathrm{max}}$, where 
$\tilde{\sigma}_{\mathrm{max}}$ is the maxium of 
$\tilde{\sigma}_{\mathrm{reg}} (\omega)$. The lines mark the positions of $\omega_p$ 
determined from $L=20$ ED data, Fig.~\ref{fig1}(c), and are put also on top of $L=32$ 
MCLM data, Fig.~\ref{fig1}(d). Here, the main message is that generally $\omega_p$ 
weakly depend on $L$, whereby the exception are the regime of $\Delta \to 1$, but 
also at $0.1 < \Delta <0.5$ where we notice possibly significant reduction of 
$\omega_p$ with $L$. In order to properly resolve the latter regime $\Delta \to 0$, 
we apply also the degenerate-perturbation-theory (DPT) method \cite{mierzejewski21}. 
Results confirm a pronounced peak in 
$\tilde \sigma_{\mathrm{reg}}(\omega \sim \omega_p)$ with $\omega_p \propto \Delta$
and $\tilde \sigma_{\mathrm{reg}}(\omega \ll \omega_p) \sim \omega^2$. Furthermore, 
results for the 1D Hubbard model obtained at generic quarter-filling, $\bar n = 1/2$, 
also reveal - besides more universal structure of dynamical charge conductivity 
$\tilde \sigma_c(\omega)$ - no clear indication for finite diffusion component 
$\tilde \sigma_{c,\mathrm{reg}}^0$.
 
\section {Spin conductivity in the XXZ chain} 
We consider in more detail 1D anisotropic XXZ spin model,
\begin{equation}
H = \frac{J}{2} \sum_{i} \left( \mathrm{e}^{i\phi} S^+_{i+1} S^-_i + \mathrm{H.c.}\right)
+ J \Delta \sum_i S^z_{i+1} S^z_i, \label{xxz}
\end{equation}
on a chain with length $L$ and with generalized PBC. 
Here, $S^{\alpha}_i$ with $\alpha=+,-,z$ represent the standard $S=1/2$ operators. We 
focus on the metallic regime with the anisotropy parameter $ 0 < \Delta < 1$, 
revealing the $T>0$ dissipationless transport with $D>0$. We further on evaluate only 
canonical systems with zero magnetization, i.e., $S^z_{tot}=0$. At fixed $L$ and at 
PBC, the results might depend on the phase shift $\phi$. Since in the following we 
numerically study systems $L = 4 {\cal L}$, i.e., $L = 16, 20, \dots, 32$, we choose 
$\phi = \pi/L$ (equivalent to anti-PBC) in order to stay consistent with our previous 
studies of the fermionic version of the model, i.e., the $t$-$V$ model 
\cite{zotos96,mierzejewski21}. Note that considered systems at $S^z_{tot}=0$ have
even number of fermions. We further on use $\hbar=k_B=1$ as well as fix $J=1$ as the 
unit of energy. 

We concentrate on high-$T$ dynamical spin conductivity 
$\tilde \sigma(\omega)=T \sigma(\omega)$, within the linear response theory for 
$T \gg J$ given by
\begin{equation}
\tilde \sigma(\omega)= \frac{\pi}{L N_{st} } \sum_{n,m} |\langle n|j| m\rangle |^2 \delta(\omega - \epsilon_m+\epsilon_n),
\label{sig}
\end{equation}
expressed here in terms of many-body (MB) eigenstates $|n\rangle $ and eigenvalues 
$\epsilon_n$, with the spin current 
$j= (J/2) \sum_i \left( i \mathrm{e}^{i\phi} S^+_{i+1} S^-_i + \mathrm{H.c.}\right)$, and 
$N_{st}$ as the total number of MB states for given $L$ and $S^z_{tot}$. Besides 
$S^z_{tot}=0$ we use also translational symmetry of the model, \eqref{xxz}, so that 
calculation of Eq.~\eqref{sig} is performed as the sum over all wavevector-$q$ sectors. 

\subsection{Numerical method}
For smaller systems $L \leq 20$ (for the XXZ model) we evaluate Eq.~\eqref{sig} directly via 
the full ED finding all $|n\rangle, \epsilon_n$. For larger $L$ we employ the MCLM 
\cite{long03,prelovsek11}, used in several studies of dynamical transport in (mostly 
disordered) spin systems \cite{prelovsek17}. Since the aim in the present problem is to achieve 
besides large $L$ (with the Hilbert space up to $N_{st} \sim 10^8$) also well resolved spectra 
at $\omega \sim 0$, we upgrade MCLM by enabling very high frequency resolution 
$\delta \omega \sim 10^{-3} J$.
The calculation steps are the following: (a) The sum over eigenstates in Eq.~\eqref{sig} is 
replaced with the microcanonical state $|\Psi_{\cal E}\rangle$ corresponding to the energy 
${\cal E}$. The latter is obtained with $M_L \gg 1$ Lanczos steps using the 
operator $V=(H-{\cal E})^2$. For larger systems, such procedure is not expected to converge to 
exact eigenvalues of $H$, but rather to wavefunction with a small energy dispersion 
$\sigma^2_{\cal E}= \langle \Psi_{\cal E} |V| \Psi_{\cal E}\rangle$. By performing Lanczos 
procedure twice and by extracting lowest eigenfunction only, we avoid the full diagonalization 
of $M_L \times M_L$ matrix. (b) In the second step, $\tilde \sigma(\omega)$ is evaluated as the 
resolvent 
\begin{equation}
\tilde \sigma(\omega) = \frac{1}{L} \mathrm{Im} \langle \Psi_{\cal E}| j 
 \frac{ i}{\omega+i \eta + {\cal E} - H} j |\Psi_{\cal E} \rangle, \label{res}
\end{equation}
evaluated again with $M_L $ Lanczos steps starting with initial wavefunction 
$j |\Psi_{\cal E} \rangle $.
Finally, Eq.~(\ref{res}) is expressed in terms of 
continued fractions and is evaluated for small $\eta$.
$M_L$ determines the frequency resolution as $\delta \omega \sim \Delta E/M_L$, where 
$\Delta E$ is the energy span of $H$ for fixed $L$. Since for given $M_L$ we have 
$\sigma_{\cal E} < \delta \omega \sim \eta$, one can directly choose desired 
$\delta \omega$ by increasing $M_L$, even well beyond $M_L \sim 10^4$. To reduce 
statistical error, we use besides translational symmetry (with $M_q=L$ different q) 
also additional sampling over $M_s \geq 1$ targeted energies ${\cal E}_{qs}$ with a Gaussian 
distribution corresponding to high-$T$ value of $\langle H^2 \rangle \sim L/8 $ for given $L$. The final result is the average 
$\tilde \sigma(\omega) = 1/(M_qM_s) \sum_{qs} \tilde \sigma({\cal E}_{qs},\omega$).

\begin{figure}[tb]
\includegraphics[width=1.0\columnwidth]{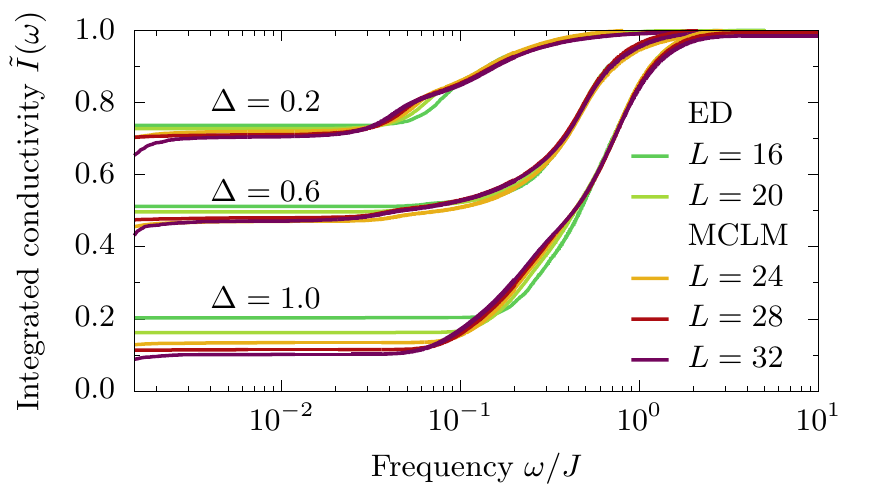}
\caption{Integrated and normalized dynamical spin conductivity 
$\tilde I(\omega)=I(\omega)/I_\infty$ in a log-$\omega$ scale within the XXZ spin 
chain for selected $\Delta = 0.2, 0.6, 1.0$, as calculated for different sizes, i.e., 
$L = 16, 20$ using full ED, and $L=24,28,32$ using MCLM.}
\label{fig2}
\end{figure}

In the present application to XXZ chain, we choose $M_L=2 \times 10^4, M_s=4$ for $L=28$ 
system and $M_L=10^4, M_s =1 $ for $L=32$ chain which has $N_{st}\sim 2 \times 10^7$ in 
given symmetry sector with fixed $q$ and $S^z_{tot}$. It is worth noting that that our method 
is related to dynamical-quantum-typicality 
(DQT) approach to time-dependent correlation functions at $T \gg 0$ 
\cite{steinigeweg15}. The latter method employs the evolution in the time domain, 
whereby our results would imply reaching times up to 
$\tau \sim 2\pi/\delta \omega > 5.10^3/J$ for considered systems.

To capture $\tilde \sigma(\omega)$ spectra including the singular ${\cal D} >0$ 
component, it is convenient to present the integrated intensity 
$I(\omega) = (2 T/\pi) \int_0^\omega \sigma(\omega') \mathrm{d}\omega'$. Here we note 
that $I(\omega \to 0) = 2 {\cal D}$, while the sum rule at $S^z_{tot}=0$ gives 
$I_\infty = I(\omega \to \infty)= \langle j^2 \rangle/L = (1/8) (1 + 1/(L-1))$ 
including small $1/L$ correction. To reveal the feasibility of the MCLM, in 
Fig.~\ref{fig2} we present the renormalized 
$\tilde I(\omega)=I(\omega)/I_\infty$ as obtained for different sizes $L = 16 - 32$ 
(whereby $L=16,20$ results are obtained via full ED) for three characteristic 
$\Delta = 0.2, 0.6, 1.0$. Results are presented in log-$\omega$ scale in order to 
amplify the low-$\omega$ regime at $\omega > 2.10^{-3} $, beyond the ${\cal D} >0$ 
contribution, in MCLM smeared within $\omega < \delta \omega$. It should be stressed 
that the MCLM method calculates the whole dynamical response 
$\tilde \sigma(\omega)$, so
it is essential to have high $\delta \omega$ resolution in order to well separate the 
singular ${\cal D} >0$ (restricted to $\omega < \delta \omega$) contribution from the regular 
part $\tilde \sigma_{\mathrm{reg}}(\omega)$. On the other hand, the separation of regular and
singular part is facilitated since we find for all considered (finite-size $L\leq 32$) systems 
$\tilde \sigma_{\mathrm{reg}}(\omega \to 0) \sim 0$. This seperation is additionally 
tested via comparison to the ED results, where the ballistic component is obtained directly 
from diagonal matrix elements in Eq.~(\ref{sig}).

Presented results reveal generally quite small $\tilde \sigma_{\mathrm{reg}}(\omega)$ at 
low-$\omega <0.02$ for 
presented $\Delta\lesssim 0.6$ and for all $L$. Still, $\omega$-dependence is quite 
pronounced in the intermediate frequency regime, depending crucially on $\Delta$. On 
the other hand, on approaching the isotropic case $\Delta \to 1$, the results on 
Fig.~{\ref{fig2} confirm pronounced $L$ dependence of the dissipationless component 
$I(\omega \to 0) =2{\cal D}$. The latter is expected to vanish for 
$L \to \infty $ \cite{prelovsek04} for $\Delta = 1$, while the regular part should 
approach the superdiffusive transport \cite{ilievski18,ljubotina19,gopalakrishnan19,denardis19,denardis21}.

\subsection{Commensurate anisotropies $\Delta_m$}

Let us first focus on results for the commensurate values of the anisotropy 
$\Delta_m =\cos(\pi/m) $ and on the possibility of the coexistence of ballistic 
component ${\cal D} >0$ and finite diffusion, i.e., 
$\tilde{\sigma}_{\mathrm{reg}}^0 > 0$. Since $L$ dependence might be important, in 
Fig.~\ref{fig3} we present results obtained for different $L=20 - 32$ for two 
specific commensurate points $m=3$ ($\Delta_3 =0.5$) and $m=4$ 
($\Delta_4 = 1/\sqrt{2} \simeq 0.707$). The most clear case appears to be 
$\Delta=0.5$, where taking into account finite 
smearing due to $\delta \omega$, it is hard to claim finite 
$\tilde \sigma^0_{\mathrm{reg}} > 0$. Results are more consistent with the 
presence of a soft gap for $\omega < \omega_g$.
The gap may be roughly estimated from the change of slope of 
$\tilde \sigma_{\mathrm{reg}}(\omega)$. In particular, for $\Delta=\Delta_3$ and $L=32$ such 
change of slope is well visible in Fig.~\ref{fig3}(a) at $\omega_g\simeq 0.08$.
The regular part appears to vanish inside the gap as 
$\tilde \sigma_{\mathrm{reg}}(\omega<\omega_g) \propto \omega^\zeta$
with $\zeta >1$ as shown in the inset in Fig. \ref{fig3}. However, $\omega_g$ reveals some
$L$ dependence which could be possibly made compatible even with its vanishing in the
thermodynamic limit. 
At least, we can put some upper bound on
$\tilde \sigma_{\mathrm{reg}}^0$. Given that the optical conductivity increases with the frequency for 
\mbox{$0<\omega<0.4$} and that $\tilde \sigma_{\mathrm{reg}}(\omega>0.15)$ shows no 
finite-size effects, we estimate 
$\tilde \sigma^0_{\mathrm{reg}} < \sigma_{\mathrm{reg}}(\omega=0.15)<0.1$.
This bound can be compared with the value 
$\tilde \sigma^0_{\mathrm{reg}} \sim 0.0685$ obtained by the GHD approach 
\cite{denardis18,agrawal20}. On 
Fig.~\ref{fig3}(a) we plot also the result obtained with tDMRG on much bigger 
system \cite{karrasch15}, but with a restricted time span $\tau < 35/J$. The 
agreement for larger $\omega>0.1$ is quite satisfactory indicating less relevant $L$ 
dependence in this regime. On the other hand, the deviation for $\omega <0.1$ is not 
surprising since (referring to the authors of \cite{karrasch15}) the spectra for $\omega<0.15$ 
are beyond the reach of their study. 

Our results for $\Delta_4$ are somewhat less conclusive due to quite pronounced 
low-$\omega$ peak at $\omega \sim 0.1$ which, however, does not shift 
significantly with $L$ (in contrast to the case $\Delta \to 1$ as presented in 
Fig.~\ref{fig2}). We should also point out that when speculating on possible closing 
of the (again soft) spectral gap in Fig.~\ref{fig3}(a) or Fig.~\ref{fig3}(b) 
for both $\Delta_m$ with $L \to \infty$, this should be done by keeping moments 
$\mu_n=\int_0^\infty\,\omega^n \tilde \sigma(\omega)\mathrm{d}\omega$ unchanged, 
since they are correctly reproduced in finite systems (as well as in MCLM) up to to 
high order $n = L$. In any case, our results for $m=4$ should be compared with 
the GHD result $\tilde \sigma^0_{\mathrm{reg}} \sim 0.14$ \cite{denardis18a}. 

\begin{figure}[tb]
\includegraphics[width=1.0\columnwidth]{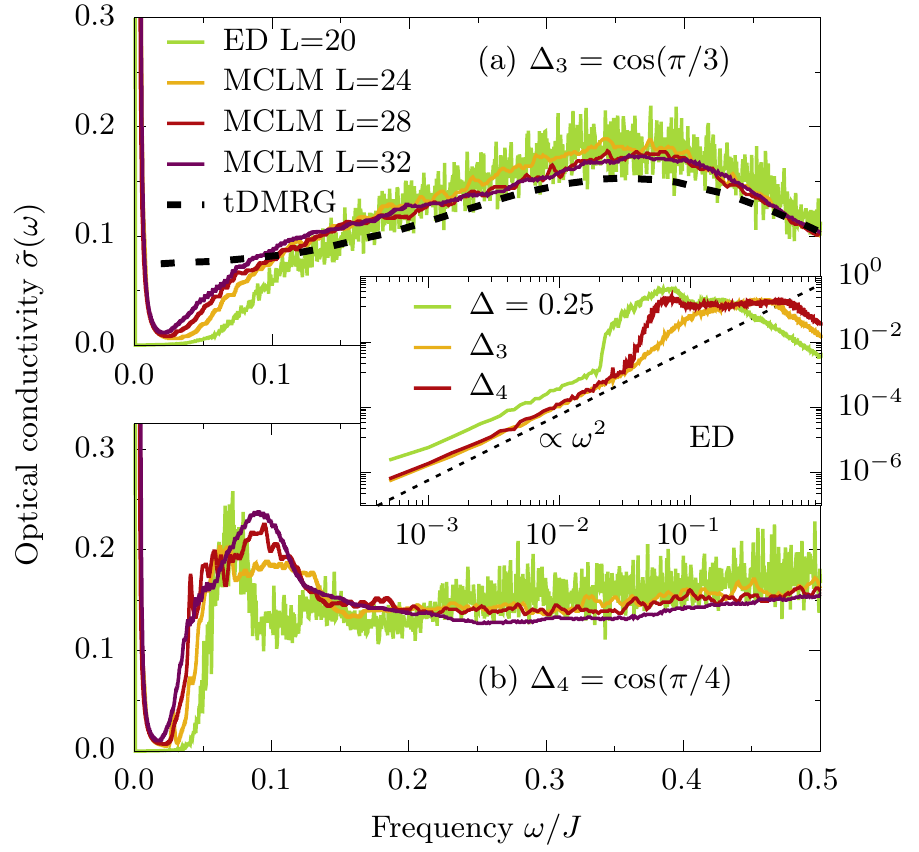}
\caption{ Spin conductivity $\tilde \sigma(\omega)$ obtained for different 
$L = 20 - 32$ for two commensurate $\Delta = \cos(\pi/m) $: (a) $m=3$ 
($\Delta =0.5$), where the dotted line is the result of tDMRG \cite{karrasch15}, and 
(b): $m=4$, $\Delta = 1/\sqrt{2}\simeq 0.707$. Inset shows $\tilde \sigma_{\rm reg}(\omega)$ 
(using  logarithmic scale) obtained from ED for selected values of $\Delta$ together
with a guideline $\sim \omega^2$.}
\label{fig3}
\end{figure}

\subsection{General structure of $\tilde{\sigma}(\omega)$}
\label{sec:stru}

To discuss the structure of $\tilde \sigma(\omega)$ in the whole regime $\Delta <1$, 
we first present in Fig.~\ref{fig4} the evolution of $I(\omega)$ with $\Delta$ as 
obtained via MCLM on $L=32$ system with the resolution 
$\delta \omega \sim 10^{-3} $. The advantage of $I(\omega)$ is that it yields direct 
information on the stiffness $2 {\cal D}=I(\omega \to 0)$ discussed before 
\cite{bertini21,mierzejewski21}). Results indicate that in the regime 
$\Delta \leq 0.5$ the major part of the $I(\omega)$ response is in the 
dissipationless component ${\cal D}$. Moreover, our results for ${\cal D}$ are in 
this regime quantitatively consistent with previous DQT analysis 
\cite{steinigeweg14,steinigeweg15} and with the DPT method for $\Delta \to 0$ 
\cite{mierzejewski21}, which both yield a value significantly above the one 
representing the lower-bound/GHD result \cite{prosen13,bertini21,bulchandani21}.
For $\Delta \geq 0.5$ the major part in $I(\omega)$ is in the regular response and the 
mismatch with GHD lower bound is less evident. 
It follows from Fig.~\ref{fig2} that ED/MCLM results for $\Delta \to 1$ exhibit considerable 
$L$ dependence showing up in the apparent ${\cal D} >0$, but also in closing of the 
finite-size gap (being $\omega_g \sim 0.05$ for $L=32$). It is, however, well visible 
that for $\omega> \omega_g$ at $\Delta =1$ the spectra in Fig.~\ref{fig4} can be well 
fitted with $I(\omega) \propto \omega^{2/3}$ implying the superdiffusive response 
$\tilde \sigma(\omega) \propto \omega^{-1/3}$ 
\cite{ilievski18,ljubotina19,denardis19,gopalakrishnan19,agrawal20,denardis21,bulchandani21}.

\begin{figure}[tb]
\includegraphics[width=1.0\columnwidth]{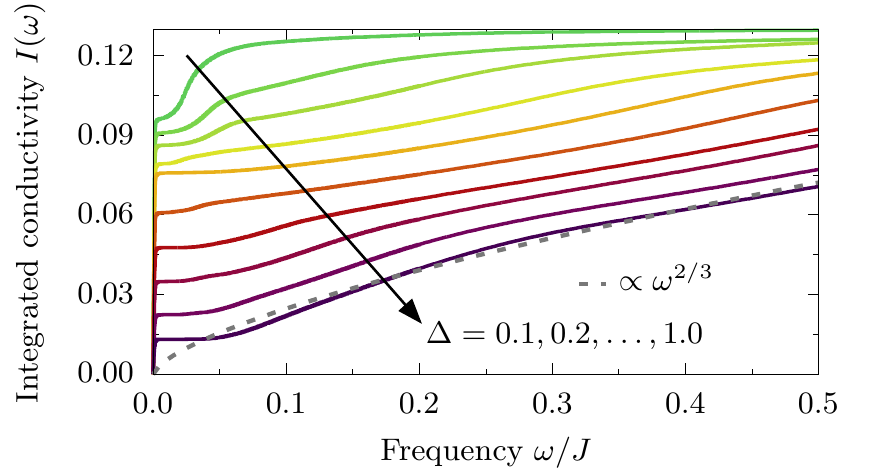}
\caption{ Integrated spin conductivity $I(\omega)$ for the whole range of 
$\Delta = 0.1 - 1.0$, as obtained via MCLM on $L=32$ XXZ spin chain. For $\Delta=1$ 
we show also the fit for superdiffusive $I(\omega) \propto \omega^{2/3}$.}
\label{fig4}
\end{figure}
 
Beyond dissipationless ${\cal D}>0$ component, $I(\omega)$ in Fig.~\ref{fig4} reveals 
quite complex evolution of $\tilde \sigma_{\mathrm{reg}}(\omega)$ with $\Delta$, in 
particular for $\Delta > 0.2$ there are visible more than one inflexion points with 
$d^2 I(\omega)/d\omega^2=0$ corresponding to peaks in 
$\tilde \sigma_{\mathrm{reg}}(\omega)$. 
Moreover, a direct information on $\tilde{\sigma}_{\mathrm{reg}} (\omega)$ is presented in 
Fig.~\ref{fig1} (obtained on $L=20$ via ED and $L=32$ via MCLM). Note that in the 
latter case, at $\omega<\delta \omega \sim 10^{-3}$, the spectra are dominated by 
the singular (but broadened) dissipationless component ${\cal D}>0$. Nevertheless, 
the structure of peaks consistent with Fig.~\ref{fig1}(a) is still visible in Fig.~\ref{fig1}(b). As 
discussed recently \cite{mierzejewski21}, at small $\Delta < 0.2$ a single peak at 
$\omega_p \propto \Delta$ dominates $\tilde \sigma_{\mathrm{reg}}(\omega)$ and is analyzed in Sec.~\ref{sec:delta0}
in more detail. 

In the intermediate regime $0.2 \leq \Delta < 0.5$ the structure of $\sigma_{\mathrm{reg}}(\omega)$ evolves into two 
peaks, i.e., the upper part still with $\omega_{p1} \propto \Delta$ retaining most of 
the sum rule of $\sigma_{\mathrm{reg}}(\omega)$, while the lower peak $\omega_{p2}$ 
splits off and is expected to vanish at commensurate $\Delta_3 =0.5$ as 
$\omega_{p2} \propto |\Delta_3 - \Delta| $. Such evolution is consistent with the 
specific behavior at commensurate points, $\Delta_m = \cos(\pi/m)$ 
\cite{zotos99,prosen11,prosen13}, where also additional degeneracies (diagonal as 
well as off-diagonal in $S^z_{tot}$ \cite{zadnik16,mierzejewski21}) exist. The latter 
macroscopic degeneracies of the energy spectrum should coincide with discontinuous 
variation (jumps) of the stiffness ${\cal D}$ at these particular values 
$\Delta = \Delta_m$. At the same time, the frequency moments $\mu_n$ (up to large 
$n \sim L$) are continuous functions of $\Delta$, even close to 
commensurate values. The simplest scenario which may capture both features is that 
the missing spectral weight, 
${\cal D}(\Delta_m)-\lim_{\Delta \to \Delta_m} {\cal D}(\Delta)$, is transferred to 
narrow low-frequency peaks centered at frequencies 
$\omega_{pm} \propto |\Delta_{m+1} - \Delta|$. In particular, we can confirm 
such a scenario by our detailed numerical analysis (ED for $L=20$) for the vicinity of 
$\Delta_4=1/\sqrt{2}$, as presented in Fig.~\ref{fig5}(a). It clearly shows the 
emergence of symmetric peaks at $\Delta \sim \Delta_4$. 
At the same time, also the widths of these peaks appear to scale 
as $\delta \omega_{pm} \propto |\Delta_{m+1} - \Delta|$. Both these facts 
show qualitative analogy to the situation close to 
$\Delta=\Delta_2 = 0$, presented in Fig.~\ref{fig5}(b), where the development 
with a single peak at the frequency $\omega_p \propto \Delta$ can can be followed much more in detail 
 \cite{mierzejewski21} (see also more elaborate analysis and discussion in 
Sec.~\ref{sec:delta0}). It should be however, noted that in a finite system, one can 
follow a jump in ${\cal D}$ and related emergence of peaks in 
$\tilde \sigma_{\mathrm{reg}}(\omega)$ only provided that the particular MB states at 
given $L$ and $S^z_{tot}$ display a large degeneracy at commensurate 
$\Delta_m$. This is indeed the case for $\Delta_2$ and $\Delta_4$ assuming the considered here finite 
systems $L=4 {\cal L}$. On the other hand, at $\Delta_3=0.5$ our systems do not 
reveal explicit (large) degeneracy, so the above phenomena cannot be followed in the 
very vicinity of $\Delta_3$, but they become more evident with the increasing system size 
as in Figs.~\ref{fig1}(b) and (d) for $L=32$. 

\begin{figure}[tb]
\includegraphics[width=1.0\columnwidth]{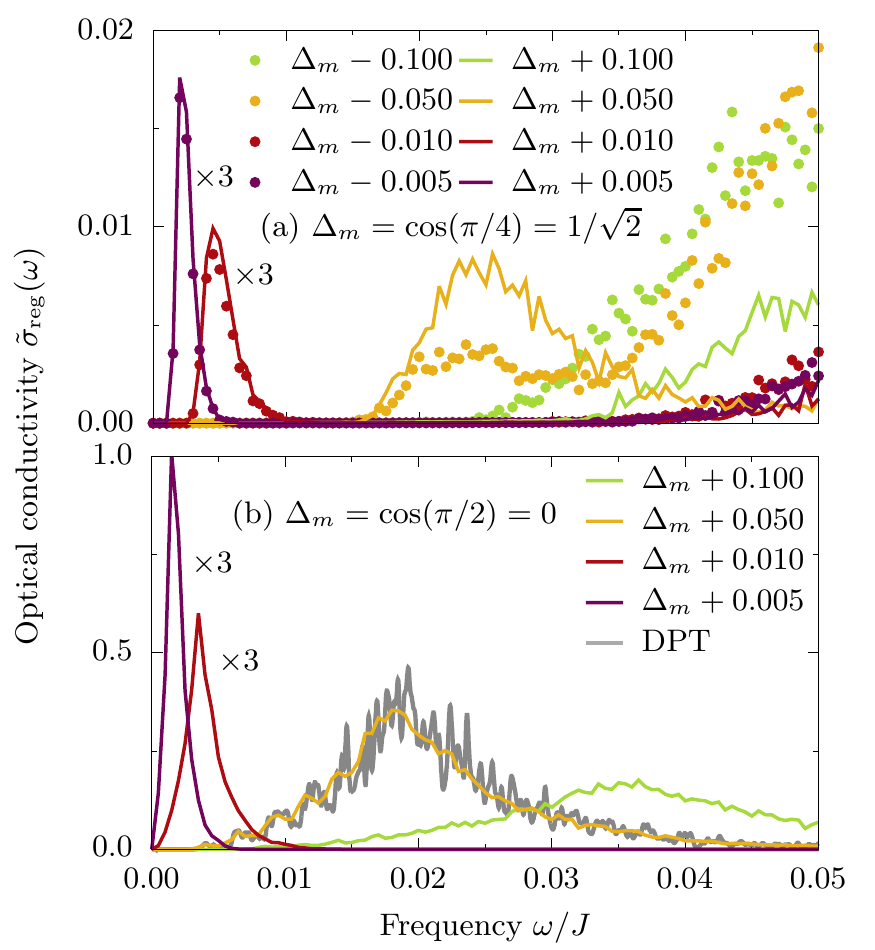}
\caption{Detailed analysis of $\tilde{\sigma}_{\mathrm{reg}}(\omega)$ spectra in 
vicinity of (a) $\Delta_4=1/\sqrt{2}$ and (b) $\Delta_2=0$ anisotropy, as calculated 
for $L=20$ sites by ED. The DPT result is also marked in panel (b).} 
\label{fig5}
\end{figure}

The above scenario should imply also for $\Delta > \Delta_{3} = 0.707$ further peaks 
emerging and leading to quite complex low-$\omega$ structure of 
$\tilde \sigma_{\mathrm{reg}}(\omega)$ on approaching $\Delta \to 1$. However, 
results for $\Delta > 0.8$ should be taken with some reservation, since they already 
reveal significant $L$-dependence. In particular, at $\Delta \to 1$ one expects the 
vanishing coherent part ${\cal D} \to 0$, while the observed spectral gap 
$\omega_g \propto 1/L^\zeta$, well visible in Fig.~\ref{fig1}, 
below quite 
featureless spectra is known to be finite-size effect \cite{prelovsek04}. This gap reflects
the anomalous $L$-dependence of $\tilde \sigma_{\mathrm{reg}}(\omega)$ in the 
integrable XXZ chain in the insulating regime $\Delta > 1$ \cite{znidaric11} 
and, in particular, the superdiffusion at $\Delta =1$ 
\cite{ilievski18,ljubotina19,gopalakrishnan19,denardis19,denardis21}, 
discussed in connection with the Fig.~\ref{fig4}.
\subsection{ The case of small $\Delta \ll 0.5 $}
\label{sec:delta0}

Results for $L=20$ and $L=32$ shown in Figs.~\ref{fig1}(c,d) demonstrate that the 
position of the single peak at $\omega_p$ strongly depends on $L$ only for 
$\Delta < 0.5$. In this subsection we carry out the finite-size scaling of 
$\sigma_{\mathrm{reg}}(\omega \to 0)$ in the regime of small $\Delta$. At 
$\Delta = 0 $, the current operator commutes with the Hamiltonian, thus 
$\tilde{\sigma}(\omega)=2 \pi {\cal D} \delta(\omega)$ and the regular part is 
absent. There is also a large jump (discontinuity) of the stiffness at $\Delta \to 0$ 
\cite{bertini21,mierzejewski21}. Similarly to other commensurate points, the 
spectral weight ${\cal D}(0)-\lim_{\Delta \to 0} {\cal D}(\Delta)$ is transferred to 
$\omega>0$ for $\Delta \ne 0$. The latter weight forms a peak at 
$\omega_p \propto \Delta$, see Fig.~\ref{fig1} and Fig.~\ref{fig5}(b). As a 
consequence, the regime of weak interaction is unique in that the structure of 
$\tilde{\sigma}(\omega)$ is relatively simple.

\begin{figure}[tb]
\includegraphics[width=0.933\columnwidth]{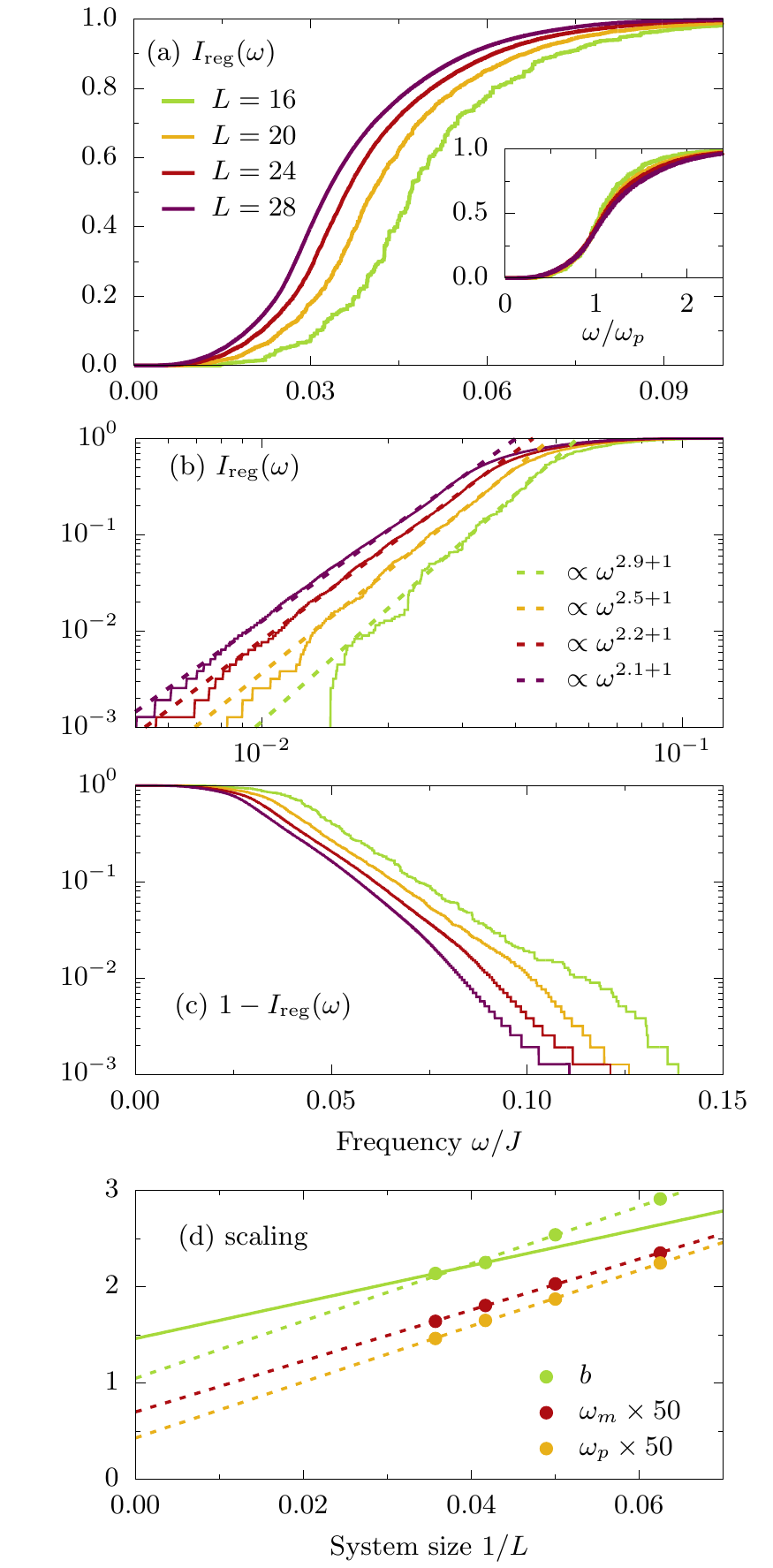}
\caption{(a) Integrated and normalized regular part of the optical conductivity 
$I_{\mathrm{reg}}(\omega)$ obtained for $L=16 - 28$ sites via the DPT calculation for 
$\Delta \to 0$. The inset shows rescaled data, $I_{\mathrm{reg}}(\omega/\omega_p)$,
where $\omega_p$ is the position of maximum, (b) The low-frequency part of 
$I_{\mathrm{reg}}(\omega \ll \Delta)$ 
fitted by $I_{\mathrm{reg}} \propto \omega^{b+1}$ (fits are shown as dashed lines). 
(c) Exponential decay of $1-I_{\mathrm{reg}}(\omega)$. (d) Finite-size scaling of the 
exponent $b$ from panel (b), where dashed line shows fits linear in $1/L$ and the 
continuous guideline goes through results for $L=24$ and $L=28$. In (d) we present 
also the position of peak $\omega_p$ and median $\omega_m$ together with the linear 
fits.}
\label{fig6}
\end{figure}

To evaluate $\tilde{\sigma}(\omega)$, we employ a recent approach which targets the 
regime of $\Delta \ll 0.5$. For the details we refer to \cite{mierzejewski21}, where 
the focus was on the stiffness ${\cal D}$, which requires calculation of the diagonal 
matrix elements $\langle n|j| n \rangle$ for $\Delta \ne 0$. The eigenstates 
$| n\rangle$ can be obtained via the DPT starting from the noninteracting case, 
$\Delta=0$, with a highly degenerate energy spectrum. The value of ${\cal D}$ 
obtained from the latter approach agrees with the ED results up to the numerical 
precision \cite{mierzejewski21}. We extend these calculations to obtain also the 
off-diagonal matrix elements, $\langle n|j| m \rangle$, and the corresponding 
$\tilde \sigma_{\mathrm{reg}}(\omega)$ defined in Eq.~(\ref{sig}). In 
Fig.~\ref{fig5}(b) we compare the DPT and ED results for $L=20$ sites and find 
perfect agreement. Note however, that the DPT does not require any 
$\omega$-smoothing, i.e., it has full frequency resolution, and allows for full 
analysis of large system sizes, i.e., up to $L=28$.

Fig.~\ref{fig6}(a) shows the integrated regular part $I_{\mathrm{reg}}(\omega)$ for (
arbitrarily chosen small) $\Delta=0.1$ and various system sizes $L$. For convenience, 
the presented quantity is normalized so that $I_{\mathrm{reg}}(\omega\to \infty)=1$. 
Since we apply the perturbative approach, the very same results hold true for 
arbitrary $\Delta \ll 0.5$, up to rescaling of frequency, 
$\omega \to \omega\Delta/0.1$. In order to estimate the 
$\tilde \sigma_{\mathrm{reg}}(\omega \to 0)$ in the $L \to \infty$ limit, we first 
note the power-law dependence of the \mbox{low-$\omega$} part, 
$I_{\mathrm{reg}}(\omega \ll \omega_p)\propto \omega^{b+1} $ so that 
$\tilde \sigma_{\mathrm{reg}}(\omega) \propto \omega^b$. The power-law fits are shown as 
straight lines in Fig.~\ref{fig6}(b) and the resulting exponents, $b$, are plotted in 
Fig.~\ref{fig6}(d) as a function of $1/L$. The latter function shows a positive 
curvature, thus a straight (continuous, green) line going through the results for two 
largest $L$ may serve as a lower bound for $b$. These results indicate that 
$1 \le b \le 2$ for $L \to \infty$ with the possibility of the analytic form 
$\tilde \sigma_{\mathrm{reg}}(\omega\ll \omega_p)\propto \omega^2$ \cite{herbrych12}. 
Fig.~\ref{fig6}(c) shows the asymptotic behavior of $1-I_{\mathrm{reg}}(\omega)$, 
where one observes that the response decays exponentially for $\omega \gg \omega_p $. 
Finally, we estimate the $L$-dependence of $\omega_p$, which corresponds to the 
position of largest slope in Fig.~\ref{fig6}(a), 
and is shown in Fig.~\ref{fig6}(d). For comparison, we show also the median value, 
$\omega_{m}$, defined as $I_{\mathrm{reg}}(\omega_m)=0.5$.
Both quantities apparently follow linear dependence in $1/L$. The extrapolation 
(which within DPT represents more a lower bound) of this trend suggests that in the 
thermodynamic limit $\omega_{m}>0$ and $\omega_p>0$. Since DPT results 
in Fig.~\ref{fig6}(d) suggest that $\omega_{m} \sim \omega_p$ for all $L$, we
present in the inset of Fig.~\ref{fig6}(a) also the \sout{renormalized} rescaled quantity,
$I_{\mathrm{reg}}(\omega/\omega_p)$, which indeed appears to be quite universal for all $L$. 
Still, this is only approximately true, since also exponent $b$ changes (slightly) with $L$,
as summarized also in Fig.~\ref{fig6}(b,d).

\section{1D Hubbard model}
Another relevant model for the possible coexistence of ballistic and diffusive 
transport is the integrable 1D Hubbard model of interacting fermions,
\begin{equation}
H= - t \sum_{i s} (c^\dagger_{i+1,s} c_{i,s} + \mathrm{H.c.}) + U \sum_i n_{i \uparrow} n_{i \downarrow},
\label{hub}
\end{equation}
again on a chain of length $L$ with PBC, where we consider states with fixed number 
$N_\uparrow, N_\downarrow$ of up- and down-spin fermions, respectively. The 
properties of the model depend on the filling (density) 
$\bar n = (N_\uparrow + N_\downarrow)/L$ and the magnetization 
$\bar m = (N_\uparrow - N_\downarrow)/L$. Here, one can define both charge $j_c$ and 
spin $j_s$ currents, so we can discuss corresponding conductivities as well as 
stiffnesses ${\cal D}_c, {\cal D}_s$, respectively. Most recent studies of 1D Hubbard 
model focused on the half-filling case, i.e., $\bar n = 1$ and $\bar m =0$ 
\cite{bertini21, ilievski17, ilievski171,karrasch17}, where both stiffnesses vanish, 
i.e., ${\cal D}_c={\cal D}_s=0$, due to the relation with the isotropic Heisenberg 
model. This is not the case for $\bar n \neq 1$ \cite{ilievski171}, where 
${\cal D}_c$ and ${\cal D}_s$ also exhibit in general a jump in the proximity to the 
noninteracting limit $U \to 0$ \cite{mierzejewski21} (in analogy to the 
$\Delta \to 0$ in the XXZ model). It should be also reminded that the first example 
of the ballistic transport at $T>0$ was the Hubbard chain with one particle 
$N_\uparrow =1$ in a bath of fermions corresponding to $N_\downarrow \sim L/2$ 
\cite{castella95,castella96}.

As in the XXZ spin chain, in the Hubbard model at $U>0$ there might persist finite 
dc contributions $\tilde \sigma_c(\omega \to 0) >0, \tilde \sigma_s(\omega\to0) >0$ 
\cite{ilievski171} besides the ballistic components at general 
$\bar n \neq 1, \bar m \neq 0$. On the other hand, the evolution with $U>0$ is 
expected to be more generic since, unlike the XXZ chain, there are no anomalies 
associated with particular $U$ values. We numerically investigate here only the case 
of charge conductivity $\tilde \sigma_c(\omega)$ at quarter filling $\bar n=1/2$ and 
$\bar m=0$. In Fig.~\ref{fig7} we present results obtained on the chain of $L=20$ 
sites via MCLM, presented in Fig.~\ref{fig7}(a) as a scan through the range of smal/modest $0 < U/t \leq 2$ and in Fig.~\ref{fig7}(b) for selected modest/large 
$U/t =1,2,4,8$. It should be first mentioned, that quite similar results emerge when 
performing ED for $L=16$, hence the $L$-dependence appears to be weak, at least in 
the accessible $L$-range.

Apart from the dissipationless component, which in the presented quarter-filling case 
takes approximately half of the sum rule \cite{mierzejewski21}, the variation for 
small/modest $U/t >0$ is quite analogous to the $\Delta \gtrsim 0$ in the XXZ spin 
chain. For $U/t \leq 1.5$ the spectra $\tilde \sigma_{\mathrm{reg}}(\omega)$ are 
dominated by a single peak at $\omega_p \propto U$ with a vanishing dc limit 
$\tilde \sigma^0_{\mathrm{reg}} \to 0$. For larger $U/t > 1.5$ the structure in Fig.~\ref{fig4} develops more 
components: a) large-$\omega$ contribution which directly reflects the scale 
$\omega \sim U$, b) the remaining low-$\omega$ structure apparently still reveals two 
not well separated peaks, where the lower one remains at $\omega_{p1} \sim t/2$ and 
the upper broader one $\omega \lesssim 3t/2$ is related to the incoherent bandwidth. 
Most important for the present study, presented results suggest vanishing 
$\tilde \sigma_{\mathrm{reg}}(\omega \to 0)$ or at least allow only for a very small upper 
bound of $\tilde{\sigma}^0_{c,\mathrm{reg}}$.

\begin{figure}[tb]
\includegraphics[width=1.0\columnwidth]{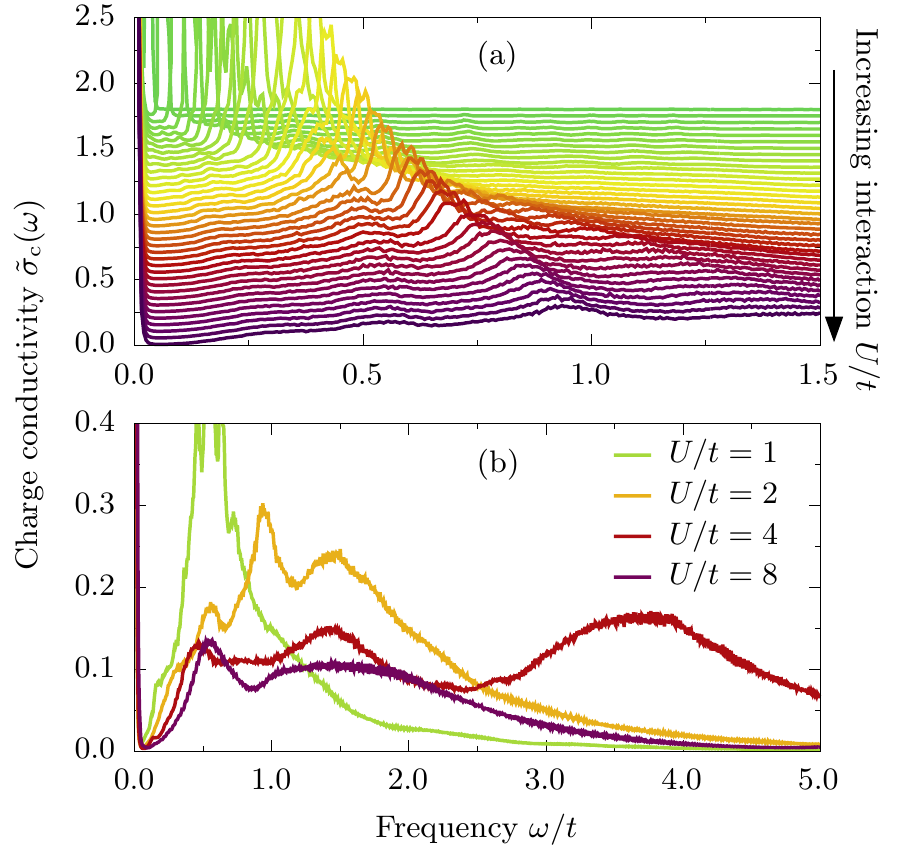}
\caption{(a) Charge conductivity $\tilde \sigma_{c}(\omega)$ for the 1D Hubbard model 
at quarter filling $\bar n=1/2$ and zero magnetization $\bar m =0$, obtained via MCLM 
for a chain of $L=20$ sites and $M_L=5\cdot10^3$ steps. (a) The scan for the whole 
interaction $0 < U/t <2$ range. (b) Spectra for selected moderate/large
 $U/t=1, 2, 4, 8$.}
\label{fig7}
\end{figure}

\section{Conclusions}
In this paper we presented the numerical results for dynamical transport response, 
i.e., high-temperature $T\to \infty$ conductivities $\sigma(\omega)$, in two 
integrable lattice 1D models, i.e., the anisotropic Heisenberg XXZ model and the 
Hubbard model. Since our aim was to resolve the low-$\omega$ behavior in terms of 
dissipationless ${\cal D}$ component and the potential remnant diffusive contribution 
$\sigma_{\mathrm{reg}}(\omega \to 0)$, we adapted the MCLM approach by allowing for 
large systems (up to $L=32$ for the XXZ and $L=20$ for the Hubbard model), as well as 
high frequency resolution $\delta \omega/(J,t) \sim 10^{-3}$. Our results can be 
summarized as:

\noindent a) Concerning the comparison with previous numerical results (nearly 
exclusively) on the XXZ chain, we are in agreement with DQT results for the stiffness 
${\cal D}(\Delta) = I(\omega \to 0)/2$ obtained for system sizes $L \leq 30$ with PBC 
\cite{steinigeweg14,steinigeweg15}. Apparently, we also agree with the results of 
tDMRG method \cite{karrasch15,karrasch17} obtained for larger systems $L \leq 200$, 
e.g., presented for $\Delta=0.5$ in Fig.~\ref{fig3}b. Since the latter time-dependent 
correlations are followed only up to times $\tau J \sim 35$, one cannot uniquely 
extract results for $\omega/J < 0.15$ \cite{karrasch15}, which is the essential range 
for the understanding of dynamics in the metallic $\Delta<1$ regime. 

\noindent b) Besides well established ballistic ${\cal D}>0$ contribution, regular 
part of dynamical spectra $\tilde \sigma(\omega)$ are quite complex in the XXZ model, 
revealing several-peak structure at general $\Delta \neq \Delta_m$. The most 
controlled regime appears to be that of small $\Delta \ll 0.5$, where both ED and 
MCLM calculation at finite $\Delta$ confirm single-peak structure of 
$\tilde \sigma(\omega)$ with the position at $\omega_p \propto \Delta$ and fast 
exponential-like decay for $\omega> \omega_p$. The latter is also consistent with the 
calculation within the DPT method for $\Delta \to 0$, performed here exactly up to 
system sizes of $L=28$. Our results reveal 
$\tilde\sigma_{\mathrm{reg}} (\omega< \omega_p) \propto \omega^b$ with $ 1\le b\le 2$ 
and $ \omega_p>0$ for $L \to \infty$, i.e., absence of a diffusive contribution and a 
qualitative agreement with the argument following from the level-crossing scenario 
\cite{herbrych12}. However, finite-size effects are still substantial so it is hard 
to exclude that $\omega_p=0$ for $L \to \infty$. 
 
\noindent c) Quite informative are the results for commensurate 
$\Delta=\Delta_m$, in particular for $\Delta_3 = 1/2$ and somewhat less for 
$\Delta_4=1/\sqrt{2}$, where Fig.~\ref{fig3} does not leave much room for dc 
diffusion $\tilde \sigma^0_{\mathrm{reg}}>0$ in our finite-size systems, although
the observed finite-size gap $\omega_g$ reveals some $L$ dependence, 
so that we cannot exclude its closing for $L \to \infty$ and possible agreement 
with the GHD results for $\tilde \sigma^0_{\mathrm{reg}}$ \cite{denardis18,agrawal20}. 

\noindent d) The spectral evolution is most difficult to follow in the vicinity of 
commensurate anisotropy $\Delta \sim \Delta_m$, which becomes very involved in the 
regime $\Delta > 0.5$. Namely, each $\Delta_m$ appears to be a source for 
additional structure appearing as the peak at 
$\omega_{pm} \propto |\Delta - \Delta_{m+1}|$. Such development can be, e.g., 
directly followed via ED and MCLM for $m=2$ and $m=4$, emerging from 
the lifting of additional large (exponentially increasing with $L$) degeneracies at $\Delta_m$
(see Fig.~\ref{fig1} and Fig.~\ref{fig5}). Less conclusive is the case of 
$\Delta_3 = 0.5$, where large-$L$ results still reveal such peaks, but considered PBC 
systems with $L = 4{\cal L}$ do not seem to exhibit explicitly such 
degeneracies. In any case, such scenario should effectively reappear with increasing 
$L$, as we also confirm by comparing in Fig.~\ref{fig1} results for systems with 
$L=20$ and $L=32$.

\noindent e) It should be remarked that the emerging spectral components for 
$\Delta \neq \Delta_m$ are also the origin for the GHD expectation concerning 
a singular $\tilde \sigma_{\mathrm{reg}}(\omega) \propto \omega^{-\alpha}$ with 
$\alpha>0$ \cite{agrawal20} in the vicinity of commensurate $\Delta_m$. To justify 
the latter, besides the jump of ${\cal D}$ and of the remaining sum rule 
$\tilde \sigma_{\mathrm{reg}}(\omega)$, an additional assumption is a power-law decay in time
of current correlation $\langle j(\tau)j \rangle - \langle j(\infty)j \rangle$. 
However, our finite-$L$ results for $\tilde \sigma_{\mathrm{reg}}(\omega)$ 
are compatible with an oscillating time-evolution to 
final $\langle j(\tau \to \infty)j \rangle \propto {\cal D} >0$ value, which is 
visible also in the tDMRG results in \cite{karrasch15}.

\noindent f) The evolution within the 1D Hubbard model appears somewhat simpler when 
considering, e.g., the regular part of the high-$T$ charge conductivity, 
$\tilde \sigma_{c,\mathrm{reg}}(\omega)$, in the ballistic regime away from 
half-filling $\bar n = 1$. For the particular case of quarter-filling and zero 
magnetization ($\bar n=1/2$, $\bar m=0$), our results reveal a single-peak structure 
for modest $U/t<1$ with the peak $\omega_p \propto U$ and vanishing 
$\tilde \sigma^0_{c,\mathrm{reg}}$, quite in analogy with the $\Delta \to 0$ in the 
XXZ chain. For larger $U/t > 1$, a large-$\omega$ peak splits off with 
$\omega_{p} \sim U$ while $\omega <2t$ regime still reveals some nontrivial two-peak 
structure, but again apparently with no diffusive contribution at low-$\omega$. The 
latter is consistent with previous results for particular case of the 1D Hubbard 
model, i.e., representing a single particle in a fermionic bath \cite{castella96}.
 
\noindent g) The arguments and predictions for the possible coexistence of ballistic 
and diffusive transport emerge within the GHD approach, which directly implies the 
limit $L\to \infty$. Since we hardly see clear evidence for 
$\tilde \sigma_{\mathrm{reg}}^0 >0$, even in largest systems ($L=32$), a minimum 
conclusion could be that such a diffusion is anomalous, i.e., is not reflected in a 
physically-relevant mean free path $\lambda < L$. Such case is not excluded and can 
emerge also in integrable quantum lattice systems. Closely related example is the 
high-$T$ diffusion in the easy-axis XXZ model at $\Delta \geq 1$, where ${\cal D}=0$ 
and apparent $\tilde{\sigma}^0_\mathrm{reg} >0 $ implies an effective 
$\lambda_{eff} \sim 1$, whereas $\tilde \sigma(\omega) $ still exhibits anomalous 
$\omega \propto 1/L$ finite-size effects 
\cite{prelovsek04,steinigeweg12,bertini21}. In fact, we can confirm such anomalous 
$L$-dependence in our analysis on approaching $\Delta \to 1$, where we confirm the 
superdiffusive dynamical scaling $\tilde \sigma(\omega) \propto \omega^{-1/3}$ 
\cite{ilievski18,ljubotina19,denardis19,agrawal20,bulchandani21}. 
Such quite explicit $L$-dependence signals the importance of the 
relevant order of limits $L \to \infty$ and $ t \to \infty$ (or $\omega \to 0$).
Namely, different quantities can require different limits, and in particular
$\sigma^0_{\mathrm{reg}}$ (evaluated at $L \to \infty$ first) might not correspond to energy 
dissipation (heating) inside the system, as found, e.g., for the easy-axis side $\Delta>1$ \cite{mierzejewski11},
i.e., the diffusion (as e.g. evaluated within GHD) might be dissipationless. 

\noindent h) Finally, it should be reminded that even a weak integrability-breaking perturbation (e.g., a 
single impurity in the XXZ chain \cite{barisic09}) in the finite MB system with PBC 
turns, e.g., the singular $\sigma(\omega)$ response into a Lorentzian-type normal 
diffusion with well defined characteristic $\lambda$. 

\begin{acknowledgments}
The authors thank T. Prosen and E. Ilievski for fruitful discussions. P.P. 
acknowledges the support by the project N1-0088 of the Slovenian Research Agency. 
M.M. acknowledges the support by the National Science Centre, Poland via project 
2020/37/B/ST3/00020. J.H. acknowledges the support by the Polish National Agency of 
Academic Exchange (NAWA) under contract PPN/PPO/2018/1/00035. The numerical 
calculation were partly carried out at the facilities of the Wroclaw Centre for 
Networking and Supercomputing.
\end{acknowledgments}

\bibliography{manusigint}
\end{document}